\documentclass{article}
\begin{document}
\pagenumbering{arabic}

\title{Nontopological structures in the baby-Skyrme model}
\author{B.M.A.G. Piette\thanks{e-mail address: B.M.A.G.Piette@durham.ac.uk},\\
W.J. Zakrzewski \thanks{e-mail address: W.J.Zakrzewski@durham.ac.uk} \\
Department of Mathematical Sciences\\University of Durham\\ Durham DH1 3LE,
UK}
\maketitle

\begin{abstract}
We report our observations that the baby-Skyrme model in (2+1) dimensions
possesses  non-topological stationary solutions which we call psedo-breathers.  
We discuss their properties and present our results on their 
interaction with the topological skyrmions.

\end{abstract}

\def\p#1{\partial_#1}
\def\mod#1{ \vert #1 \vert }

\newcommand{\ee}{\end{equation}}
\newcommand{\be}{\begin{equation}}
\newcommand{\Ref}[1]{(\ref{#1})}

\section{Introduction.}

The $(2+1)$-dimensional baby-Skyrme field theory model is described by 
the Lagrangian density
 
\be 
L=F_{\pi}\bigl({{1\over2}\partial_{\alpha}\vec \phi \partial^{\alpha} 
\vec \phi - 
{k^{2}\over4}
(\partial_ {\alpha} \vec \phi \times \partial_{\beta}\vec \phi )
(\partial^{\alpha} \vec \phi \times \partial\sp{\beta} \vec \phi)-
 \mu^{2}(1-\vec n \cdot\vec \phi)\bigr)}.
\label{eLagPhi}
\ee

 Here $\vec \phi \equiv (\phi_{1},\phi_{2},\phi_{3}) $ denotes a triplet 
of scalar real 
fields which satisfy the constraint ${\vec \phi}^{2}=1$;
$( \partial_{\alpha} \partial^{\alpha}=\partial_{t}\partial^{t}-\partial_{i} 
\partial^{i})$.
As mentioned in \cite{RBBSKTwo}-\cite{rTigran} the first term in \Ref{eLagPhi} is 
the familiar Lagrangian density of the pure
$S\sp2$ $\sigma$ model. The second term, 
fourth order in derivatives, is the (2+1) dimensional
analogue of the Skyrme-term of the 
three-dimensional Skyrme-model \cite{rSkyrme}.
The last term is often referred to  as 
a potential term.  
The last two terms in the Lagrangian \Ref{eLagPhi}
are added to guarantee the stability of the skyrmion\cite{rLPZ}.

This model has static solutions which describe topologically stable
field configurations called
skyrmions\cite{rBBSKOne}. Such solutions have been studied in detail; 
results can be found
in \cite{rBBSKOne} and \cite{RBBSKTwo}. 
However, this model also has many other interesting solutions. In particular,
it has plane wave like solutions. Moreover, any solution of the Sine-Gordon
model is also a solution of the baby-Skyrme model.
To see this observe that if we parametrise   $\vec \phi$ as
 
\be    
\vec\phi=(\sin{f}\cos{\psi},\sin{f}\sin{\psi},\cos{f}),  
\label{eVecPhi}
\ee 
and then assume that $\psi$ is constant 
the equation of motion for the field $f$ reduces to
\be
\partial_\mu \partial^\mu f+ \mu^{2}\sin(f)=0,
\label{eSineGordon}
\ee
which, of course, is the sine-Gordon equation. 
Thus all solutions of this equation, kinks, breathers {\it etc}
are automatically solutions of the baby-Skyrme model.

However, as we have shown in \cite{newSasha} all such solutions are unstable.
Any small perturbation destroys them.
For any of these solutions we have found that as soon as some region of the 
wave is perturbed, the wave collapses around this point, emitting radiation. 
The collapse front then propagates rapidly along the solitonic wave 
destroying it completely.

While performing simulations with large amplitude breather waves we have
observed the formation of a radially symmetric breather-like soliton. 
As this solution seemed very similar to the pulsons observed by
\cite{rChrLom} we have decided
to look at it in more detail.
\section{Pseudo-breathers}
Looking at the time evolution of the field configurations seen in our
simulations has lead us to make the following ansatz for our 
pseudo-breather fields:
\be    
\vec\phi=(\sin{f(r,t)},0,\cos{f(r,t)}).  
\label{eVecBreather}
\ee
Then, simple calculations show that the
 equation of motion reduces to the radial sine-Gordon equation:
\be
f_{tt} - f_{rr} - {f_{r}\over r} + \mu^{2}\sin(f)=0.
\label{eSinGordon}
\ee

 This equation has already been studied in \cite{rChrLom}, where it
was shown that it had time dependent solutions similar to a breather, but
which radiate their energy and slowly die out. The authors 
of  \cite{rChrLom}  called 
such configurations  pulsons.
 
In \cite{rTDsinG} we have also shown 
that there exist stable time
dependent solutions of \Ref{eSinGordon}.
 Radial field configurations of \Ref{eSinGordon} radiate
relatively quickly when their amplitude of oscillation is relatively small, 
{\it ie} when the value of $f$ never becomes larger than $\pi/2$ at the origin 
(these are the pulsons studied in \cite{rChrLom}.) 
However, when the amplitude of oscillation is larger than  $\pi/2$ the 
general field
configuration  radiates its energy very slowly  asymptotically reducing 
the amplitude of 
oscillation to a value a little larger than $\pi/2$  and settles at a pseudo-breather with
 a period of oscillation $T \sim 20.5$ 
(when $\mu^2= 0.1$).  By trial and error we have found that

\be
f(r,0) = 4\, atan(C \,exp\bigl(-{2\over\pi}{\mu r\over K} atan({\mu r\over K})\bigr) )
\label{ePBinit}
\ee
and $
{\partial f \over \partial t} (r,0) = 0$
with $K = 10$ and $C= \tan(\pi/8)$ is a good initial condition which leads to
 this metastable pseudo-breather solution.

We have checked that this pseudo-breather solution is stable when enbedded
into 
the $S^2 $ model. 

Its energy  is given by
$$
  E_{PB}\, \sim \,3.97
$$
which shows that it is approximately $2.5$ heavier than the baby-skyrmion.
Moreover, its topological charge density is identically zero 
thus it has enough energy
to decay into a skyrmion anti-skyrmion pair. However, we have not seen
 such a decay in our simulations.

As our pseudo-breather can only be determined numerically, in practice
we have to use the field configurations which we have found in our
numerical simulations. 
 The excess of energy over the final 
configuration can then be seen as an excitation energy which is slowly
radiated away.
The scattering properties of pseudo-breathers are quite interesting.
When the pseudo-breathers are embedded into the baby-skyrmion model 
the field configurations have an extra degree of freedom 
corresponding to their orientation in the $\phi_1 , \phi_2$ plane.
When two pseudo-breathers are set at rest near each other,
the force between them depends  on their relative orientation: when 
they are parallel to each other and oscillate in phase, they
attract each other, overlap and form a new structure which appears to be
an exited pseudo-breather. This pseudo-breather then 
slowly radiates away its energy. The non-topological nature of 
pseudo-breathers allows them to merge and form a new structure
of the same type.

If the two pseudo-breathers are anti parallel, {\it ie} if they oscillate
completely out of phase, then the force between them is repulsive. When the
two pseudo-breathers have a different orientation they slowly rotate
themselves until they become parallel; then they move towards each 
other and form an exited pseudo-breather structure.

When two  pseudo-breathers are sent towards each other with some kinetic
energy, the scattering is more complicated. Depending on the initial speed or
the scattering impact parameter, they either merge into a single 
pseudo-breather or they undergo a forward scattering. The details 
of these scattering properties are given in \cite{rTDsinG}.

When a skyrmion and a pseudo-breather are put at rest next to each other
the overall interaction between them  makes the skyrmion slowly move
away from the pseudo-breather 
while the pseudo-breather looses some of its
energy faster than when it is placed there by itself.

To scatter a skyrmion with a pseudo-breather we have placed the 
pseudo-breather soliton at rest, 
and we have sent the skyrmion towards it. We have performed 
this scattering for different orientations of the pseudo-breather, 
for different values of the impact parameter and 
for  different speeds.

The results of our simulations are reported in \cite{newSasha}. We have
found that the amount of energy lost by the pseudo-breather during the scattering is
larger when  the overlap between the skyrmion and the 
pseudo-breather, both in time and space, is greater. 
In some cases the pseudo-breather is completely destroyed by the collision. 
The oscillation of the pseudo-breather makes the interaction time dependent
and, as a result,  the pattern of scattering angles 
observed in the simulation is quite complicated. 

\section{Conclusions}
We have shown that the baby-Skyrme model, in addition 
to skyrmions  has 
other solutions which are nontopological in nature
 and periodic in time.
 These new solutions  have interesting scattering properties with each other and with skyrmions. They are stable but  can be destroyed if 
 perturbed too much.

\section{Acknowledgments}

We would like to thank A. Kudryavtsev for useful discussions.

%\refout 
 
\end{document}